\documentstyle[a4,12pt,epsfig]{report}

\begin{document}

\centerline{Physics Reports {\bf 376} (2003) 1-103}
\bigskip
\bigskip
\bigskip

\centerline{\Large\bf Magnetic light scattering in low-dimensional }
\centerline{\Large\bf quantum spin systems} 
\bigskip

\centerline{\large\bf P. Lemmens$^a$, G. G\"untherodt$^b$ and C. Gros$^c$}
\medskip

\centerline{\it $^a$ Max Planck Institute for Solid State Research, 
Heisenbergstrasse 1, D-70569, Stuttgart, Germany}

\centerline{\it $^b$ 2. Physikalisches Institut, RWTH Aachen, 
         D-56056, Aachen, Germany}

\centerline{\it $^c$ Institut f\"ur Theoretische Physik, Univ. des Saarlands, 
D-66041, Saarbrücken, Germany}

\bigskip
\bigskip
\bigskip

\hrule
\medskip

\noindent
{\bf Abstract}
\medskip

An overview of one- and two-dimensional quantum spin systems based 
on transition-metal oxides and halides of current
interest is given, such as spin-Peierls, spin-dimer, 
geometrically frustrated and ladder systems. The most significant and
outstanding contributions of magnetic light scattering to 
the understanding of these materials are discussed and compared
to results of other spectroscopies and thermodynamic measurements. 

\medskip
\noindent
{\bf Author Keywords:} Transition metal compounds; 
Low-dimensional quantum spin; Magnetic light scattering 

\medskip
\noindent
{\bf PACS classification codes:} 75.10.Jm; 78.30.-j 

\medskip
\hrule

\newpage
\tableofcontents
\setcounter{chapter}{-1}
\chapter{Preface}

This article reviews recent progress in magnetic light scattering
in one- and two-di\-men\-sional quantum spin systems. These
systems received considerable interest from both theoretical and
experimental points of view. Following the investigations of the
two-dimensional superconducting cuprates and the search for
related transition metal oxides a fascinating field of copper
oxide compounds, vanadates, manganites and nickelates opened up.
These compounds show effects of strong electronic correlations and
in particular magnetism in low dimensions.

The theory of magnetism in one dimension, on the other hand, has a
history reaching back to the origin of quantum mechanics. This is
due to the fact that a spin chain allows more easily analytical or
numerical solutions. It was found that the suppression of
``trivial" long-range order sets the stage for an enormous
complexity of possible ground states, exotic quasiparticles and
many-body states. Understanding these effects is the most
intriguing challenge at present.

A central concept in describing these low-dimensional quantum spin
systems is that of a spin liquid. This ground state is dominated
by strong quantum fluctuations, pronounced spin-spin correlations
and a suppression of long-range magnetic order. The Heisenberg
chain with isotropic-antiferro\-magnetically coupled spins (s=1/2)
represents such a state in the sense that the spin-spin
correlations decay algebraically. It is therefore often denoted as
a critical spin liquid. An interesting situation occurs when
transitions lead to a sudden change of the excitation spectrum,
e.g., the opening of an excitation gap or the formation of
long-range magnetic order. These quantum phase transitions are
driven or controlled by the exchange coupling parameters, the
exchange topology or by spin vacancies. The excitation gap may be
realized with or without a spontaneously broken translational
symmetry.

The spin-Peierls transition and the related charge ordering instability
discovered in the inorganic compounds $\rm CuGeO_3$ and $\rm NaV_{2}O_{5}$,
respectively, represent the case of broken translational symmetry. These
compounds allow to investigate the excitation spectrum going from a
homogeneous gapless to a dimerized state just as a function of temperature.
In the two-leg spin ladder system $\rm SrCu_{2}O_3$ and the chain/ladder
system $\rm Sr_{14}Cu_{24}O_{41}$ an excitation gap is realized without
breaking translational symmetry. These compounds are discussed as model
systems for an electronic mechanism of high temperature superconductivity.
The steady improvement of understanding also leads to surprising
reinterpretations of compounds that have been investigated for years. The
formerly canonical example of a spin ladder, the vanadium compound $\rm
(VO)_2P_2O_7,$ is now recognized as a spin chain with strongly alternating
coupling constants. This result has profound consequences for the
interpretation of its low energy excitations. Very important compounds that
bridge one and two dimensions and still do not show long-range magnetic
order are the spin frustrated system $\rm SrCu_2(BO_3)_2$ and the
1/5-depleted square lattice system $\rm CaV_4O_9$.

Light scattering experiments or other spectroscopic methods like
inelastic neutron scattering have been used to investigate both
the above cited and many more compounds. One of the most
significant aspects of light scattering experiments is the
observation of magnetic singlet bound states. These states
originate from strong triplet-triplet interaction and characterize
the excitation spectrum of the spin system. Recent theoretical
pro\-gress has enabled a more detailed understanding of these
effects. Parameters like dimerization, frustration, interchain
coupling, and spin-phonon coupling have an important impact on the
ground state and the excitations of a quantum spin system.

This review is organized as follows: After a brief description of the
excitations and the phase diagram of quantum spin systems given in
Chapter 1, important low-dimensional spin systems, recent
experimental results and their interpretations are discussed in
Chapter 2. Up to now no comprehensive review on this rapidly
growing field exists that also considers material aspects. Therefore we try
to balance between well established results and very recent developments. In
Chapter 3 magnetic light scattering in low-dimensional spin
systems is reviewed. The following Chapters 4 and
5 discuss magnetic bound states and quasielastic scattering.
Chapter 6 finally sums up some aspects of the present
knowledge in this field and gives an outlook to future developments.

This review benefited from discussions and collaborations with many
colleagues, too numerous to mention here completely.

We would like to thank especially E.Ya.~Sherman and R. Valent{\'{\i}} for
carefully reading the manuscript. PL and GG would like to express their
gratitude towards the recent and previous members of the light scattering
group in Aachen, e.g., K.Y.~Choi, J.~Pommer, A. Ionescu, M.~Fischer,
M.~Grove, G.~Els, and P.H.M.~van~Loosdrecht.

Further valuable discussions with F.~Mila, W.~Brenig, K. Ueda, G.S.~Uhrig,
C.~Pinettes, M.~Udagawa, V.~Gnezdilov and Yu.~G.~Pashkevich are gratefully
acknowledged. We are obliged to C. Geibel, F. Steglich, H. Kageyama, M.
Isobe, Yu. Ueda, J. Akimitsu, H. Tanaka, M. Johnsson, P. Millet, A.
Revcolevschi, B.C. Sales, W. Assmus, S. Barilo and their collaborators for
providing numerous samples. Furthermore, PL acknowledges gratefully a
stipend of the Venture Business Laboratory for a visit at the Institute of
Integrated Arts and Sciences (IIAS) Hiroshima University.

\setcounter{page}{3}
\chapter{Excitations in low-dimensional spin systems}

In strongly correlated electron systems with integer number of
electrons per site the low energy excitations are usually given by
the spin degrees of freedom. This situation is properly described
by the Heisenberg exchange spin Hamiltonian. If, in addition, the
exchange is restricted to low dimensions, then spin chains, spin
ladders, and respective systems with a more complex exchange
geometry are realized. These systems exhibit a number of unusual
properties which are related to strong quantum fluctuations. These
properties will be addressed in the following.

\begin{figure}[t]
\begin{center}
\centerline{\epsfig{figure=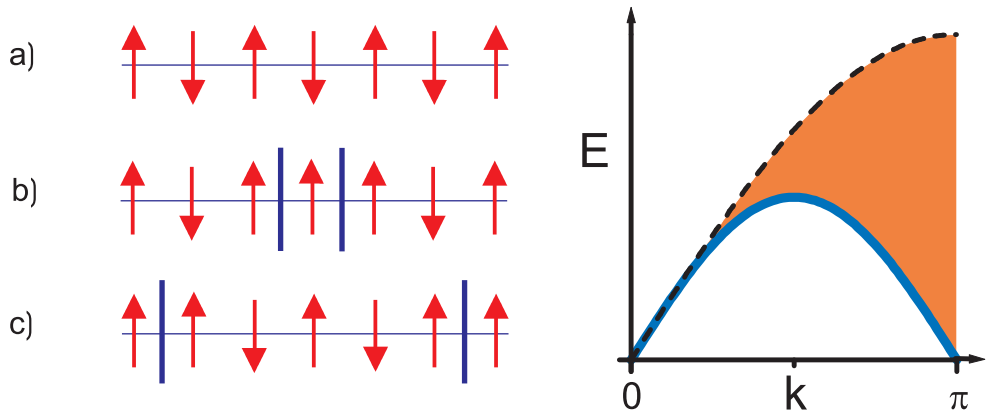,width=13.0cm}}
\end{center}
\vspace{-0.7cm}\caption[Spin excitations on a homogeneous
chain] {\label{excit-homo}Spin excitations on a homogeneous chain.
On the left hand side following a)~-~c) the generation of two
spinons (vertical bars) by a spin flip and consecutive exchange
processes is given. On the right hand side the corresponding
two-spinon continuum is shown. The spectral weight of the
continuum is maximal on the dispersing lower branch.}
\end{figure}

One-dimensional s=1/2 spin systems (spin chains) with uniform nearest 
neighbor exchange coupling show according to the Lieb-Schultz-Mattis 
theorem a degeneracy of the singlet ground state with triplet excitations. 
Assuming negligible spin anisotropies even for T=0 the 
ground state is gapless and not magnetically ordered. 
It is described by the Bethe Ansatz.
The spin-spin correlations are algebraically 
decaying typical for a quantum critical state. Triplet excitations in this 
system are not described as magnons (bosons) but as massless domain 
wall-like s=1/2 spinons (fermions). These spinons are created as pairs, 
e.g., by an exchange process. Their dynamical structure factor is therefore 
given by a gapless two-particle continuum restricted by a lower and an 
upper dispersing boundary. In Fig.~\ref{excit-homo} a sketch of the spinon 
creation and the spinon continuum is given. The spectral weight of the 
continuum is dominant close to its lower boundary.

\begin{figure}[t]
\begin{center}
\centerline{\epsfig{figure=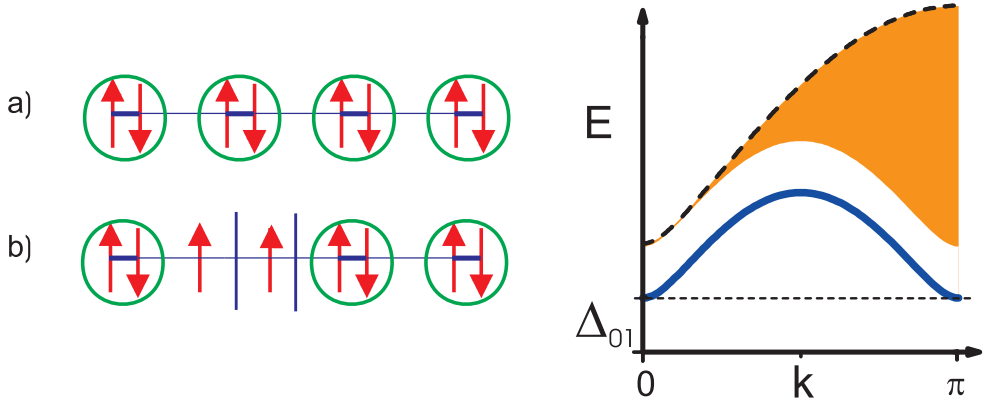,width=13.0cm}}
\end{center}
\vspace{-0.7cm}\caption[Spin excitations on a dimerized chain]{\label{excit-dimer}Sketch of spin excitations on a dimerized chain. Breaking a dimer a)~-~b) corresponds to the singlet/triplet gap $\Delta_{\rm 01}$ with the respective triplet dispersion shown on the right hand side. The continuum of ``free" triplets is reached for energies E$>$2$\Delta_{\rm 01}$. For small {\bf k} there exist a finite curvature of the dispersion
relation.}
\end{figure}

\begin{figure}[t]\begin{center}
\centerline{\epsfig{figure=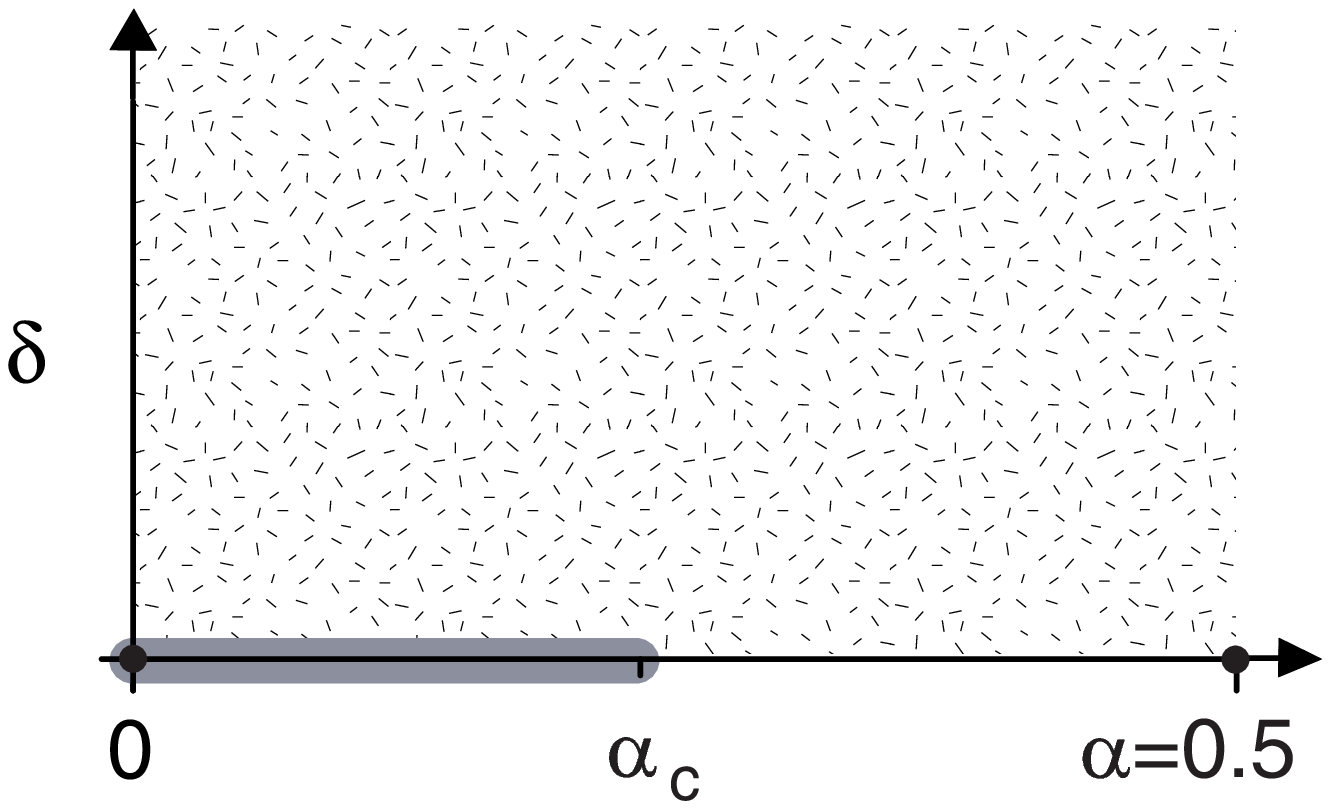,width=9.0cm}}
\end{center}
\vspace{-0.7cm}\caption[Phase diagram of dimerized and frustrated spin 
chains]{\label{diagr-ch} Phase diagram (T=0) of spin chains in dependence 
of dimerization $\delta$ and spin frustration $\alpha$. For $\delta$=0 and 
$\alpha$$\leq$$\alpha_c$=0.2412 (dark grey bar) a gapless quantum critical 
ground state exists. The remaining dashed region denotes the phase space of 
the gapped quantum disordered state. }
\end{figure}

A quantum phase transition from a gapless critical state into a
gapped state (disordered spin liquid) is induced by dimerization,
i.e. an alternation $\delta$ of the coupling constants to nearest
neighbors $\rm J_{nn}^{\pm}$=(1$\pm\delta$)$\rm J_{nn}$ along the
chain or by a sufficient frustration $\alpha$=$\rm J_{nnn}$/$\rm
J_{nn}$ due to next nearest neighbor antiferromagnetic exchange
$\rm J_{nnn}$. With dimerization the spinons are
confined into massive triplet excitations. This confinement of
spinon and antispinon composite objects (triplets) is discussed
similar to the quark confinement in particle physics.
The resulting quantum disordered
ground state is characterized by short-ranged exponential decaying
spin-spin correlations. In many cases the system is allowed to be
described as an arrangement of spin dimers. The resulting lifted
degeneracy of triplet and singlet excitations leads to an energy
gain of the system. Fig.~\ref{excit-dimer} shows a sketch of the
excitation processes in a dimerized chain with the respective
energy dispersion. The dimerization or alternation of the coupling
constants connects {\bf k}=0 and {\bf k}=$\pi$ therefore a small
part of the continuum spectral weight ($\propto$$\delta$$\rm ^2$)
is also expected for {\bf k}$\leq$$\pi$/2. In
Fig.~\ref{excit-dimer} this contribution is neglected. A (T=0)
phase diagram of dimerized and frustrated spin chains is given in
Fig.~\ref{diagr-ch}. The points ($\delta$=0,$\alpha$=0) and
($\delta$=0,$\alpha$=0.5) correspond to the Bethe Ansatz and the
Majumdar-Ghosh point. For $\delta$=0 and $\alpha_c$$<$0.3 the gap
remains numerically small.

The simplest representative of the quantum disordered state,
however, is the two-leg spin ladder with an approximately equal or
larger exchange coupling along the rungs with respect to the
coupling along the legs of the ladder. The
singlet ground state is composed of spin dimers on the rungs.
Here, the term spin liquid is even more appropriate as it is not
based on a broken translational symmetry. An excitation in this
picture of strong dimerization corresponds to the breaking of one
dimer. The energy related to this process is the singlet-triplet
gap $\rm \Delta_{01}$, see Fig.\ \ref{excit-dimer}. A coupling of
more than two chains to three-, four- or five-leg ladders leads to
the experimentally proven conjecture that ladders with an even
number of legs have a spin gap while odd-leg ladders are gapless
. ~\footnote[2]{It should be mentioned
that the combined effect of dimerization and interchain
interaction may also lead to a vanishing spin gap in a quantum
spin system. In a certain parameter space of a two-leg ladder
 with additional frustration, 4-spin cyclic
exchange or interchain interactions, quantum phase transitions to
gapless phases have been observed. On the
other hand, there is theoretical evidence for a spin gap in doped
three-leg ladders for a certain set of exchange coupling constants.}

\begin{figure}[t]
\begin{center}
\centerline{\epsfig{figure=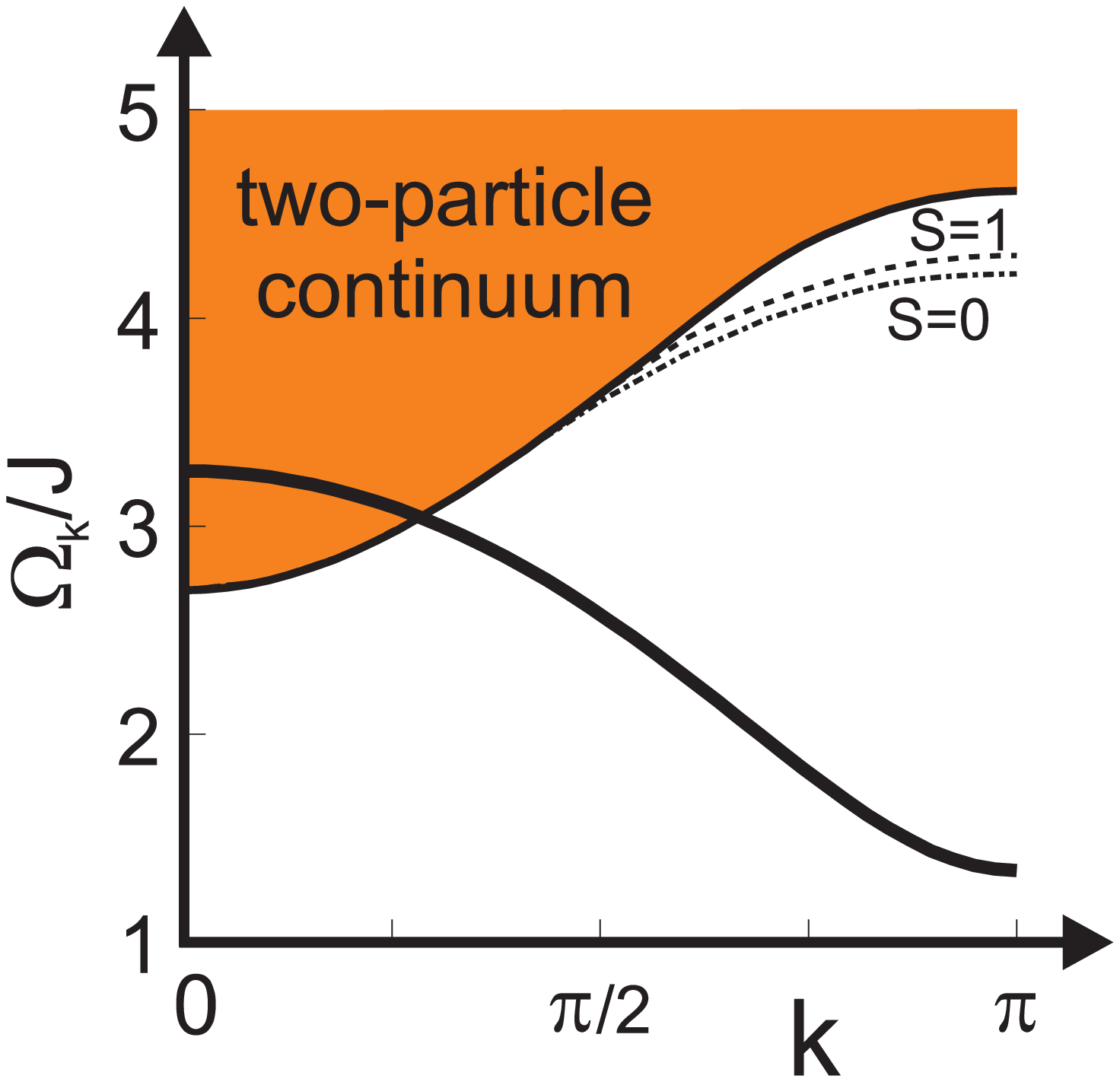,width=9.0cm}}
\end{center}
\vspace{-0.7cm}\caption[Magnetic excitations of a spin-1/2 ladder]
{\label{excit-ladder} Dispersion of magnetic excitations of a
homogeneous two-leg ladder with $J_\perp=2J_\parallel$. The
triplet branch as well as the lower boundary of the two-particle
continuum are given by full lines. The dashed lines mark singlet
(S=0) and triplet (S=1) bound states. }
\end{figure}

In the limit of a large number of coupled chains a two-dimensional
Heisenberg system is obtained and the magnitude of the respective
spin gap is going to zero. This limit may also be used to
understand the two-dimensional high temperature superconductors
(HTSC). Weakly doped two- and three-leg ladders have been
theoretically investigated in this context.
However, also in two
dimensions spin dimer ground states with a gapped excitation
spectrum are realized. This happens either if the exchange
topology is modified to favor a dimer ground state, e.g., in
removing 1/5 of the spins from a square lattice,
or due to strong frustration
(next nearest neighbor interaction).

The triplet-triplet interactions that are responsible for the
opening of the gap also lead to magnetic bound states, i.e.
triplet excitations that are bound to singlet, triplet or
quintuplet states.  The former two
states are characterized by a well-defined excitation with an
energy reduced with respect to the energy of the two-particle
continuum of ``free" triplets. If interchain or magnetoelastic
interactions are dominant bound states consist of
soliton-antisoliton pairs.
Neglecting these effects the binding energy of a bound state in a
dimerized spin chain originates from frustration. The maximum
number of bound states of a spin chain is restricted to one
singlet and one triplet state. In spin ladders with an additional
diagonal frustration the number of bound states and their binding
energy is less limited and increasing with frustration. In
Fig.~\protect\ref{excit-ladder} the excitation spectrum of a
homogeneous two-leg s=1/2 ladder is shown with a singlet and a
triplet bound state at the lower boundary of the two-particle
continuum. A quantum phase transition at a
critical frustration into a gapless phase is understood as a
condensation process of large many-particle bound states together
with a general softening of the excitation spectrum.
More generally, magnetic bound states may
therefore be used to study the triplet-triplet interaction,
determine the coupling parameters and the phase diagram of the
system.

If defects, either as localized non-magnetic vacancies or as
mobile carriers, are introduced into a quantum spin system its
excitation spectrum may change drastically.  In a 2D
square lattice doping with mobile carriers destroys long-range
N\'eel order and leads to the opening of a pseudo gap in the spin
and charge excitations. The effect of localized spin vacancies in
dimerized spin chains or in spin ladders is different. Here, a
transition from a gapped into an ordered and gapless N\'eel-type
state is induced by a seemingly negligible amount of vacancies.
This effect is based on the doping of only weakly bound spinons by
every vacancy. Thereby additional excitations are introduced in
the gap corresponding to staggered moments for sites far from the
vacancy and increasing spin-spin correlations. Interchain interaction
leads to the occurrence of magnetic order at finite temperature.

The coexistence of true long-range magnetic order and dimerization is 
possible if spatial variations of the competing order parameters are taken 
into account. This means that the excitation spectrum of 
such a system has two features, the gapped triplet mode due to dimerization 
and the gapless ``spin wave mode". The emergence of 
antiferromagnetism keeps the structure and the energy scales of these modes 
essentially unchanged as the transfer of spectral weight from the gapped to 
the spin wave mode is realized with only a small reduction of the gap and 
an increase of the spin wave velocity. The latter mode is damped with a 
broadening proportional to the square of the wave vector. 
Similar arguments have recently been used to describe the interplay or 
competition of disorder-induced antiferromagnetism and superconductivity in 
Heavy Fermion compounds. Comprehensive experimental studies 
concerning the effect of spin vacancies exist for spin ladder and dimerized 
spin chain systems. Some results including light scattering data will be 
presented in the next chapter.

The effect of mobile carriers on the gapped excitation spectrum of a spin 
liquid is directly related to the problem of an electronic mechanism for 
HTSC and not yet understood completely.
In the 2D CuO$\rm _2$ square lattice 
the doped holes are believed to form self-organized slowly fluctuating 
arrays of metallic stripes in which the motion of holes shows a locally 
quasi-one-dimensional character. 
A spin gap or pseudo gap is then the result of the spatially confined 
Mott-insulating regions of the material in the proximity of the metallic 
stripes. This effect has been described by the term ``topological doping".
Corresponding theoretical studies of weakly doped two- and 
three-leg ladders confirm these ideas in the sense that a tendency toward a 
binding effect of holes either into a superconducting condensate or charge 
ordered ground state exists. The excitation 
spectrum of the latter system is of special interest as it is separated 
into a gapless Luttinger-liquid (odd channel) and an insulating gapped spin 
liquid phase (even channel). In some sense this 
spectrum represents or mimics the scenario of spin and charge separation 
discussed for HTSC. A quantum phase transition into a superconducting state 
with d-wave character has been predicted for the three-leg ladder at higher 
doping levels.

In a very simplified picture the two channels of the three-leg ladder may 
be understood as a ``plain" ladder coupled to a chain. Holes hop back and 
forth from the chain to the ladder system. In the ladder they prefer to 
form pairs minimizing the number of broken dimers. Hopping back into the 
chain system this correlation is ``partially transferred" into the 
conducting channel. The experimental part of this 
problem is far from being completely settled and therefore under progress. 
Although only few spin ladder systems are available unusual experimental 
results pointing to {\bf k}-dependent relaxation rates different from the 
undoped material exist. In the following chapter a 
thorough review of the presently known inorganic low-dimensional compounds 
will be given focusing on the questions discussed above.

\setcounter{page}{7}
\chapter{Important inorganic quantum spin systems}
\section{Structural Considerations and Important Parameters}
\section{Quasi-Zero-Dimensional Compounds}
\subsection{Dimer compounds}
\subsection{The Cu-Tetrahedra Systems $\rm Cu_2Te_2O_5X_2$}
\section{Quasi-Two-Dimensional Compounds}
\subsection{The 1/5-Depleted System $\bf CaV_4O_9$}
\subsection{The Shastry-Sutherland System $\rm SrCu_2(BO_3)_2$}
\subsection{High Temperature Superconductors}
\subsection{The diluted square lattice system $\bf K_2V_3O_8$}
\section{Low-Dimensional Cuprates: new compounds related to high-temperature
         superconductors}
\subsection{The Two-Leg Ladder $\rm SrCu_{2}O_3$}
\subsection{The Chain/Ladder System $\rm (Sr,Ca)_{14}Cu_{24}O_{41}$}
\section{Low-Dimensional Vanadates}
\subsection{The $\rm AV_2O_5$ Family of Compounds}
\subsection{The Alternating Chain System $\rm (VO)_2P_2O_7$}
\section{Low-Dimensional Halides and Pnictides}
\subsection{The Chain System $\bf KCuF_3$}
\subsection{The $\rm ACuCl_3$ Family of Compounds}
\subsection{The Haldane System $\bf CsNiCl_3$}
\subsection{$\bf Yb_4As_3$}
\section{Inorganic Systems with Spin-Peierls and related Instabilities}
\subsection{$\bf CuGeO_3$}
\subsection{$\bf NaV_{2}O_{5}$}
\section{Magnetic Parameters of Selected Spin Systems}

\setcounter{page}{60}
\chapter{Magnetic light scattering}

\section{Two-Magnon Scattering}
\section{Light scattering in high temperature superconductors}
\section{Light Scattering in quasi-one-dimensional systems}
\subsection{The Limit of Large Dimerization}
\subsection{Nonzero Temperature}
\subsection{Defect-Induced Light Scattering}
\section{Spinon Light Scattering in $\rm CuGeO_3$}
\section{Raman scattering from spin-1/2 ladders}

\setcounter{page}{67}
\chapter{Magnetic bound states}

\section{Bound states in $\rm CuGeO_3$}
\subsection{Three-magnon scattering in $\rm CuGeO_3$}
\subsection{Defect-induced bound states in $\rm CuGeO_3$}
\section{Bound States in $\rm NaV_{2}O_{5}$}
\subsection{Experimental observations in $\rm NaV_{2}O_{5}$}
\subsection{Deficiency and substitutions on the Na Site}
\subsection{Theoretical considerations}
\section{Bound states in $\rm (VO)_2P_2O_7$}
\section{Bound states in $\rm SrCu_2(BO_3)_2$}
\subsection{Effect of substitutions} 
\subsection{Polarization dependence} 

\setcounter{page}{85}
\chapter{Quasielastic scattering in low-dimensional spin systems}

\setcounter{page}{89}
\chapter{Conclusions and outlook}

\chapter*{Acknowledgement}

\end{document}